\documentclass[aps,prl,twocolumn,preprintnumbers,superscriptaddress]{revtex4-2}
\usepackage[usenames,dvipsnames,table]{xcolor}
\usepackage{graphicx,amsmath,amssymb,amsthm,multirow,array,bm,soul}
\usepackage[mathscr]{eucal}
\usepackage[bbgreekl]{mathbbol}
\usepackage{amsfonts}
\usepackage{hyperref}
\usepackage{float}
\usepackage{url}
\usepackage{physics}
\usepackage{indentfirst}
\usepackage{lineno}
\usepackage{extarrows}
\usepackage{braket}
\usepackage{enumerate}

\DeclareMathOperator{\arccosh}{arccosh}

\DeclareMathOperator{\arcsech}{arcsech}

\begin{document}

\title{Reflected multi-entropy and its holographic dual}

\author{Ma-Ke Yuan}
\email{mkyuan19@fudan.edu.cn}
\affiliation{Department of Physics and Center for Field Theory and Particle Physics, Fudan University, Shanghai 200433, China}
\author{Mingyi Li}
\email{limy22@m.fudan.edu.cn}
\affiliation{Department of Physics and Center for Field Theory and Particle Physics, Fudan University, Shanghai 200433, China}
\author{Yang Zhou}
\email{yang\_zhou@fudan.edu.cn}
\affiliation{Department of Physics and Center for Field Theory and Particle Physics, Fudan University, Shanghai 200433, China}

\begin{abstract}
We introduce a mixed-state generalization of the multi-entropy through the canonical purification, which we call reflected multi-entropy. We propose the holographic dual of this measure. For the tripartite case, a field-theoretical calculation is performed using a six-point function of twist operators at large $c$ limit. At both zero and finite temperature, the field-theoretical results match the holographic results, supporting our holographic conjecture of this new measure.
\end{abstract}

\maketitle
\subsection{Introduction}
Quantum entanglement~\cite{Horodecki:2009zz, Amico:2007ag} has recently emerged as a key tool in understanding the AdS/CFT correspondence~\cite{Maldacena:1997re, Gubser:1998bc, Witten:1998qj}, which remains one of our most successful frameworks for probing quantum gravity. Remarkably, the Ryu-Takayanagi formula~\cite{Ryu:2006bv, Ryu:2006ef, Hubeny:2007xt, Wall:2012uf} provides a means to calculate entanglement entropy (EE) in conformal field theories (CFTs) via the minimal area of a codimension-2 surface in anti-de Sitter (AdS) space, establishing a precise link between spacetime geometry and quantum entanglement. However, EE exclusively captures quantum entanglement for pure states. To extend this connection to mixed states, holographic duals for several other measures, such as the entanglement of purification~\cite{Takayanagi:2017knl, Nguyen:2017yqw, Caputa:2018xuf}, reflected entropy~\cite{Dutta:2019gen, Jeong:2019xdr, Hayden:2021gno}, logarithmic negativity~\cite{Kudler-Flam:2018qjo, Kusuki:2019zsp, Dong:2021clv, Dong:2021oad, Dong:2024gud}, and odd entanglement entropy~\cite{Tamaoka:2018ned, Mollabashi:2020ifv}, have recently been studied. Another important direction lies in the investigation of multipartite correlation measures~\cite{Walter:2016lgl, Nezami:2016zni, Zou:2020bly, Agon:2022efa, Penington:2022dhr, Hubeny:2018ijt, He:2019ttu, Hernandez-Cuenca:2023iqh}, whose holographic duals play a critical role in understanding the emergence of bulk spacetime from the boundary many-body quantum entanglement. The aim of this Letter is to bridge the gap between multipartite correlations in mixed states and spacetime geometry. 

For a given multipartite mixed state, we can canonically purify it to obtain a pure state~\cite{Dutta:2019gen}. Using the recent proposal for pure state multipartite entanglement measure~\cite{Gadde:2022cqi} based on the replica trick and permutation invariance, one can compute the so-called multi-entropy for the purified state. We propose this as an intrinsic multipartite measure for the original mixed state, which we term {\it reflected multi-entropy}. Unlike multi-entropy~\cite{Gadde:2023zzj, Gadde:2023zni, Harper:2024ker, Gadde:2024jfi}, reflected multi-entropy is intrinsically well-defined in the ultraviolet, thus circumventing the divergence issue that plagues EE as well as multi-entropy in quantum field theory. We note that while the multipartite generalizations of reflected entropy have been studied in~\cite{Bao:2019zqc, Chu:2019etd}\footnote{A related measure called the multipartite entanglement of purification has been investigated in~\cite{Umemoto:2018jpc, Bao:2018gck}. }\nocite{Umemoto:2018jpc,Bao:2018gck}, these attempts are fundamentally distinct from our approach. Crucially, unlike these previous measures in~\cite{Bao:2019zqc, Chu:2019etd}—which are defined through the von Neumann entropy on replica spaces—our proposed reflected multi-entropy is fundamentally rooted in multi-entropy, an intrinsic multipartite entanglement measure.

In the remainder of this Letter, we first introduce reflected multi-entropy as the mixed-state generalization of multi-entropy, and then propose its holographic dual. For the tripartite case, we perform a field-theoretical calculation using a six-point function of twist operators and find that it matches the holographic result, both at zero and finite temperature.
\subsection{A Generalization of multi-entropy}
\textit{Review of multi-entropy.}~Let us recall some basic notions of EE and R\'enyi entropy. Given a pure state $\ket{\psi}_{AB}$ that describes two systems $A$ and $B$, the EE between them is defined as
\begin{equation}\label{eq-EE}
S(A) = - \text{Tr}\rho_{A} \log \rho_{A}\ ,
\end{equation}
where $\rho_{A} = \text{Tr}_{B} \ket{\psi} \bra{\psi}_{AB}$ is the reduced density matrix. 
R\'enyi entropy is a one-parameter generalization of EE
\begin{equation}\label{eq-RenyiE}
    S_{n}(A) = \frac{1}{1-n}\log\text{Tr}\rho_{A}^{n}\ ,
\end{equation}
which reduces to EE~\eqref{eq-EE} in the limit $n \to 1$. This is called \textit{replica trick}. In order to introduce the multi-entropy, let us reformulate $\rho_{A}^{n}$ through the permutation and contraction of the indices of the $n$ copies of $\rho_{A}$, 
\begin{equation}
\begin{split}
\rho_{A}^{n} 
&= \left(\rho_A\right)_{\alpha_{1}}^{\alpha_{2}} \left(\rho_A\right)_{\alpha_{2}}^{\alpha_{3}} \left(\rho_A\right)_{\alpha_{3}}^{\alpha_{4}} \cdots \left(\rho_A\right)_{\alpha_{n}}^{\alpha_{1}}\\ 
&= \left(\rho_A\right)_{\alpha_{1}}^{\alpha_{\sigma\cdot 1}} \left(\rho_A\right)_{\alpha_{2}}^{\alpha_{\sigma\cdot 2}} \left(\rho_A\right)_{\alpha_{3}}^{\alpha_{\sigma\cdot 3}} \cdots \left(\rho_A\right)_{\alpha_{n}}^{\alpha_{\sigma\cdot n}}\ ,
\end{split}
\end{equation}
with $\sigma = (123\cdots n)$ the permutation acting on the index of replicas.
Guided by this key observation, in~\cite{Gadde:2022cqi} the authors defined the \textit{multi-entropy} $S^{(\mathtt{q})}$, which generalizes the concept of EE to $\mathtt{q}$-partite cases through the permutation and contraction of indices. Let us focus on tripartite pure state $\ket{\psi}_{ABC}$ to illustrate.

Like R\'enyi entropy, the entanglement measure $S_{n}^{(\mathtt{q} = 3)}$ should be determined by the choice of permutation $(\sigma_{A},\sigma_{B},\sigma_{C})$ acting on the replica indices, i.e., the way of contracting multiple density matrices.
Note that there is a equivalence relation $(\sigma_{A},\sigma_{B},\sigma_{C}) \sim (\sigma_{A},\sigma_{B},\sigma_{C})\cdot g$ allowing us to set $\sigma_{A} = \text{id}$ by choosing $g=\sigma_{A}^{-1}$, reducing the problem to $(\mathtt{q} - 1)$-parties with the entanglement measure depending solely on $(\sigma_{B},\sigma_{C})$~\footnote{In this Letter, we treat all the subsystems symmetrically.}. Both $B$ and $C$ have $n$ replicas and $(\sigma_{B},\sigma_{C})$ are chosen to be cyclic permutations of order $n$.
In total, we have $n^{\mathtt{q} - 1} = n^2$ replicas, which form a $n \times n$ square lattice. The multi-entropy is defined by choosing the replica symmetry as $\mathbb{Z}_n \otimes \mathbb{Z}_n$ with the first or second $\mathbb{Z}_n$ acting on the subregion $B$ or $C$ and cyclically permuting the replicas located in the same row or column of the lattice. 

\begin{figure}[htbp]
    \centering
    \includegraphics[width=0.9\columnwidth]{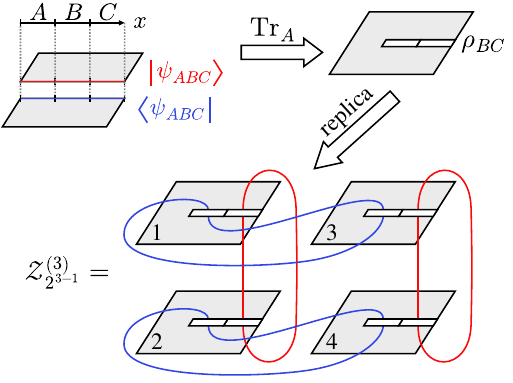}
    \caption{The construction of the $n^{\mathtt{q} - 1}$-sheet Riemann surface in the special case $\mathtt{q} = 3$ and $n = 2$. }
    \label{fig-example}
\end{figure}

For general $n$ and $\mathtt{q}$-partite entanglement, the R\'enyi multi-entropy is defined as~\cite{Gadde:2022cqi, Harper:2024ker}~\footnote{Here we follow the definition in~\cite{Harper:2024ker}. Compared with~\cite{Gadde:2022cqi}, we use a slightly different notation for the partition function on the replica space. We denote the partition function on the $n^{\mathtt{q - 1}}$-sheet Riemann surface by $\mathcal{Z}^{(\mathtt{q})}_{n^{\mathtt{q} - 1}}$ instead of $\mathcal{Z}^{(\mathtt{q})}_n$ in~\cite{Gadde:2022cqi}. }
\begin{equation}\label{eq-renyi-multiE}
S^{(\mathtt{q})}_n = \frac{1}{1 - n} \frac{1}{n^{\mathtt{q} - 2}} \log(\mathcal{Z}^{(\mathtt{q})}_{n^{\mathtt{q} - 1}}/(\mathcal{Z}_{1}^{(\mathtt{q})})^{n^{\mathtt{q}-1}})\ ,
\end{equation}
where $\mathcal{Z}^{(\mathtt{q})}_{n^{\mathtt{q} - 1}}$ is the partition function on the $n^{\mathtt{q} - 1}$-sheet Riemann surface with replica symmetry $\mathbb{Z}_n^{\otimes (\mathtt{q} - 1)}$ as described above. The multi-entropy is defined by taking the $n \to 1$ limit,
\begin{equation}\label{eq-multiE}
S^{(\mathtt{q})}= \lim_{n \to 1} S^{(\mathtt{q})}_{n}\ .
\end{equation}
For $\mathtt{q} = 3$ and $n=2$, the partition function contains four replicas with the gluing $\sigma_{B}=(13)(24)$ and $\sigma_{C}=(12)(34)$ as illustrated in Fig.~\ref{fig-example} via the Euclidean path integral~\cite{Holzhey:1994we, Calabrese:2004eu, Calabrese:2009qy}. 
This results in a contraction of four density matrices (here we abbreviate $\rho_{BC}$ as $\rho$),
\begin{equation}
\mathcal{Z}^{(3)}_{2^{3 - 1}} = \rho_{\beta_1 \chi_1}^{\beta_3 \chi_2} \rho_{\beta_2 \chi_2}^{\beta_4 \chi_1} \rho_{\beta_3 \chi_3}^{\beta_1 \chi_4} \rho_{\beta_4 \chi_4}^{\beta_2 \chi_3}\ .
\end{equation}
The corresponding R\'enyi multi-entropy is then given by \eqref{eq-renyi-multiE}
with $\mathcal{Z}_{1}^{(3)} = \rho^{\beta\chi}_{\beta\chi}$ serving as the normalization factor.

\begin{figure}[htbp]
    \centering
    \includegraphics[width=1\columnwidth]{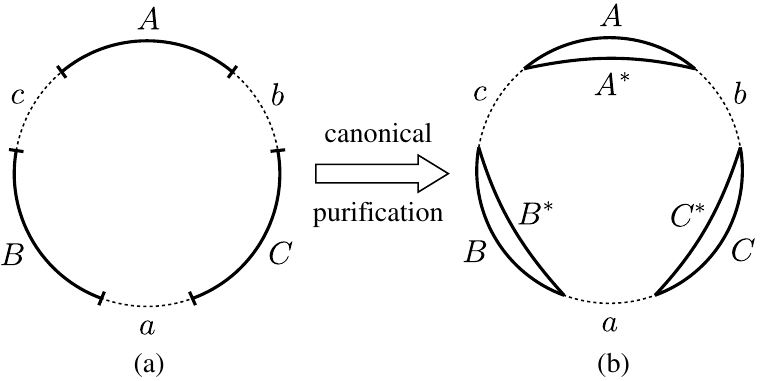}
    \caption{Canonical purification of $\rho_{ABC}$. (a) Global pure state $\psi_{AcBaCb}$ defined on a circle. Tracing out $abc$ gives a mixed state $\rho_{ABC}$. (b) Picking up another copy of $\rho_{ABC}$ and gluing these two copies along $abc$ gives a big pure state $\ket{\sqrt{\rho_{ABC}}}$~\eqref{eq-canonicalP}, which is the canonical purification of $\rho_{ABC}$. }
    \label{fig-FieldPurification}
\end{figure}
\textit{Mixed-state generalization of multi-entropy through canonical purification.}~A multipartite mixed state can be obtained by tracing out some part from a pure state. Again we will focus on tripartite case with $\mathtt{q} = 3$ but it can be generalized to larger $\mathtt{q}$ straightforwardly. Starting from a pure state $\psi_{AcBaCb}\in\mathcal{H}_{AcBaCb}$ and tracing out $abc$, we obtain a mixed state $\rho_{ABC}$, which can be canonically purified by doubling the Hilbert space (see Fig.~\ref{fig-FieldPurification} for a schematic diagram of this purification process). 
For more discussions on canonical purification we refer to~\cite{Dutta:2019gen}. The purified state can be expressed as
\begin{equation}\label{eq-canonicalP}
\begin{split}
\ket{\sqrt{\rho_{ABC}}} 
&= \ket{\sqrt{\text{Tr}_{abc}\ket{\psi_{AcBaCb}}\bra{\psi_{AcBaCb}}}}\\
&\in \left( \mathcal{H}_A \otimes \mathcal{H}_A^* \right) \otimes \left( \mathcal{H}_B \otimes \mathcal{H}_B^* \right) \otimes \left( \mathcal{H}_C \otimes \mathcal{H}_C^* \right)\ .
\end{split}
\end{equation}
We define the \textit{reflected multi-entropy} as
\begin{equation}
S_R^{(\mathtt{q} = 3)} (A; B; C) = S^{(\mathtt{q} = 3)} (AA^*; BB^*; CC^*)_{\ket{\sqrt{\rho_{ABC}}}}\ ,
\end{equation}
with $S^{(\mathtt{q})}$ the original multi-entropy~\eqref{eq-multiE}. The subscript $R$ in $S^{(\mathtt{q})}_R$ denotes ``reflected''~\footnote{The marriage with holographic purification was also mentioned in~\cite{Gadde:2023zzj} without explicit construction.}. 

Here we list some properties of reflected multi-entropy, 
\begin{enumerate}[(A)]
\item It reduces to twice the multi-entropy for pure state $\rho_{AB\cdots} = \ket{\psi}\bra{\psi}_{AB\cdots}$, 
\begin{equation}
\text{pure state:}\ S_R^{(\mathtt{q})}(A;B;\cdots) = 2 S^{(\mathtt{q})}(A;B;\cdots)\ . 
\end{equation}
\item It reduces to the reflected entropy~\cite{Dutta:2019gen} for bipartite system $\rho_{AB}$, i.e., $S_R^{(\mathtt{q} = 2)}(A;B) = S_R(A;B)$. 
\item It vanishes for factorized density matrix $\rho_{AB\cdots} = \rho_A \otimes \rho_B \otimes \cdots$, 
\begin{equation}
\text{product state:}\ S_R^{(\mathtt{q})}(A;B;\cdots) = 0\ . 
\end{equation}
However, $S_R^{(\mathtt{q})}(A;B;\cdots)$ can be nonzero for separable states. Consider a simple separable state:
\begin{equation}
	\rho_{ABC} = \frac{1}{2}\ket{000}\bra{000}+\frac{1}{2}\ket{111}\bra{111}\ .
\end{equation}
Its canonical purification is a GHZ state $\frac{1}{2}\ket{000000}+\frac{1}{2}\ket{111111}$ with nonzero $S_R^{(3)} (A; B; C) = \frac{3}{2}\log 2$. 
 
\item It is bounded from below by multi-entropy of purification,
\begin{equation}
\text{lower bound:}\ S_R^{(\mathtt{q})}(A;B;\cdots) \geq M_P(A;B;\cdots)\ ,
\end{equation}
where the multi-entropy of purification $M_P$ is defined by minimization of multi-entropy over all possible purifications. 
\item It is bounded from below by half the multipartite reflected entropy~\cite{Chu:2019etd} for holographic states,
\begin{equation}
\text{lower bound:}\ S_R^{(\mathtt{q})}(A;B;\cdots) \geq \frac{1}{2}\Delta_R(A;B;\cdots)\ . 
\end{equation}
\end{enumerate}
\subsection{Holography of reflected multi-entropy}\label{sec-holo}
\begin{figure}[htbp]
    \centering
    \includegraphics[scale=0.7]{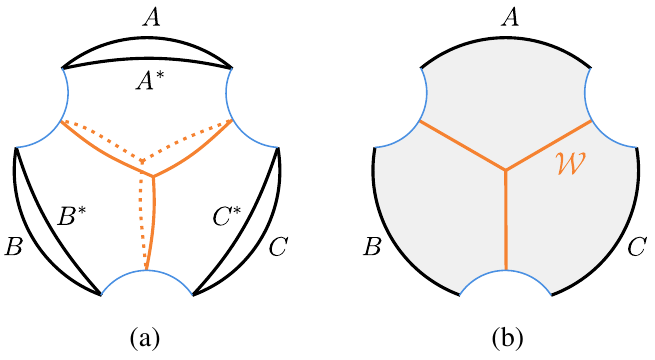}
    \caption{(a) Canonical purification of $\rho_{ABC}$ and its holographic dual. Tracing out $abc$ corresponds to gluing along the RT surface for $\rho_{ABC}$ (blue curves). The orange web $\mathcal{W}$ is the minimal web in the bulk as the holographic dual of $S^{(3)}(AA^*;BB^*;CC^*)$. (b) The single copy picture. The shadow area is the entanglement wedge of $A\cup B\cup C$ and the orange web is the holographic dual of $S_R^{(3)}(A;B;C)$. }
    \label{fig-holoPurification}
\end{figure}

Based on the canonical purification for holographic states~\cite{Dutta:2019gen} and the holographic proposal for multi-entropy~\cite{Gadde:2022cqi}, now we want to find the holographic dual for reflected multi-entropy. Consider a global pure state $\psi_{AcBaCb}$ defined on a circle as shown in Fig.~\ref{fig-FieldPurification}(a). This pure state has a classical bulk solution as its holographic dual. Now we trace out $abc$, which corresponds to retaining only the entanglement wedge for $\rho_{ABC}$ and prescinding the other portion of the bulk. To perform the canonical purification, we pick up another copy of the entanglement wedge and glue these two copies along the Ryu-Takayanagi surface (RT surface) for $\rho_{ABC}$. The resulting geometry is drawn in Fig.~\ref{fig-holoPurification}(a). 

To figure out the geometry dual of the reflected multi-entropy, which now becomes the multi-entropy between $AA^*$, $BB^*$, and $CC^*$, we need to generalize the holographic proposal for multi-entropy~\cite{Gadde:2022cqi} to multi-boundary pure states. Recall that the original holographic proposal for multi-entropy is a minimal bulk web $\mathcal{W}$ consisting of minimal codimension-2 surfaces. For a single boundary pure state, $\mathcal{W}$ is anchored at the boundaries of all subsystems $A,B,\cdots$. This condition is relaxed in the case of multi-boundary, much like the RT surfaces for multi-boundary states. Another important condition is that $\mathcal{W}$ should contain subwebs that are homologous to all the subsystems $A,B,\cdots$. This can be easily satisfied. It is then natural to propose the holographic dual of the reflected multi-entropy as the minimal surface web shown in Fig.~\ref{fig-holoPurification}(a). 
More precisely, the holographic dual of the reflected multi-entropy $S_R^{(\mathtt{q})}(A; B; \cdots)$ is the minimal surface web $\mathcal{W}$ satisfying the following topological conditions:
\begin{enumerate}[(1)]
\item $\mathcal{W}$ is anchored at the RT surface for $A\cup B\cup \cdots$. 
\item $\mathcal{W}$ contains subwebs that are homologous to all the subsystems $A,B,\cdots$. 
\end{enumerate}
The holographic reflected multi-entropy is computed by
\begin{equation}
S_R^{(\mathtt{q})} = \frac{2}{4 G_N}\min_{\mathcal{W}} L\left[\mathcal{W}\right]\ ,
\end{equation}
with $L\left[\mathcal{W}\right]$ the area of the web $\mathcal{W}$ in a single copy as illustrated in Fig.~\ref{fig-holoPurification}(b). 
\subsection{\boldmath Computation in $\text{AdS}_3/\text{CFT}_2$\unboldmath}
\begin{figure}[htbp]
    \centering
    \includegraphics[width=1\columnwidth]{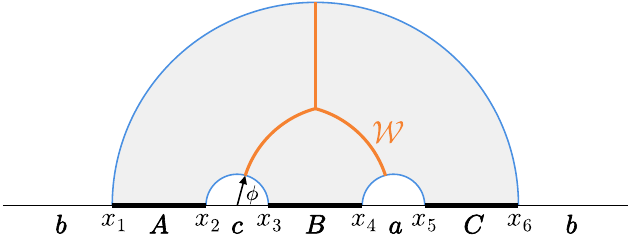}
    \caption{The holographic dual of $S^{(\mathtt{q} = 3)}_R(A;B;C)$. }
    \label{fig-conformalTrans-6pt}
\end{figure}

Let us study a simple example in AdS$_3$/CFT$_2$, as illustrated in Fig.~\ref{fig-conformalTrans-6pt}. We work in Poincar\'e half plane with the metric
\begin{equation}
\mathrm{d}s^2 = \frac{\mathrm{d}x^2 + \mathrm{d}y^2}{y^2}\ , \quad x \in \mathbb{R}\ \&\ y \in \mathbb{R}_+\ . 
\end{equation}
We consider the reflected multi-entropy $S^{(3)}_R(A;B;C)$ between subsystems $A,B,C$. $\rho_{ABC}$ is a mixed state obtained by tracing out $abc$ from the pure state $\psi_{AcBaCb}$ defined on $\mathbb{R}$ as shown in Fig.~\ref{fig-conformalTrans-6pt}. The subsystems $A,B,C$ are chosen to be $A = [x_1, x_2]$, $B = [x_3, x_4]$, $C = [x_5, x_6]$
with $x_{i = 1,\dots,6}$ parametrized by $\xi = a, b, c$,
\begin{equation}\label{eq-coorpara}
\begin{split}
x_6 &= -x_1 = b\ , \\
x_5 &= -x_2 = a+r\ , \quad x_4 = -x_3 = a-r\ .
\end{split}
\end{equation}
 In the following we will calculate $S^{(3)}_R(A;B;C)$ using both holographic method and field-theoretical method and find the agreement between them. 

\begin{figure*}[htbp]
	\centering
	\includegraphics[width=2.07\columnwidth]{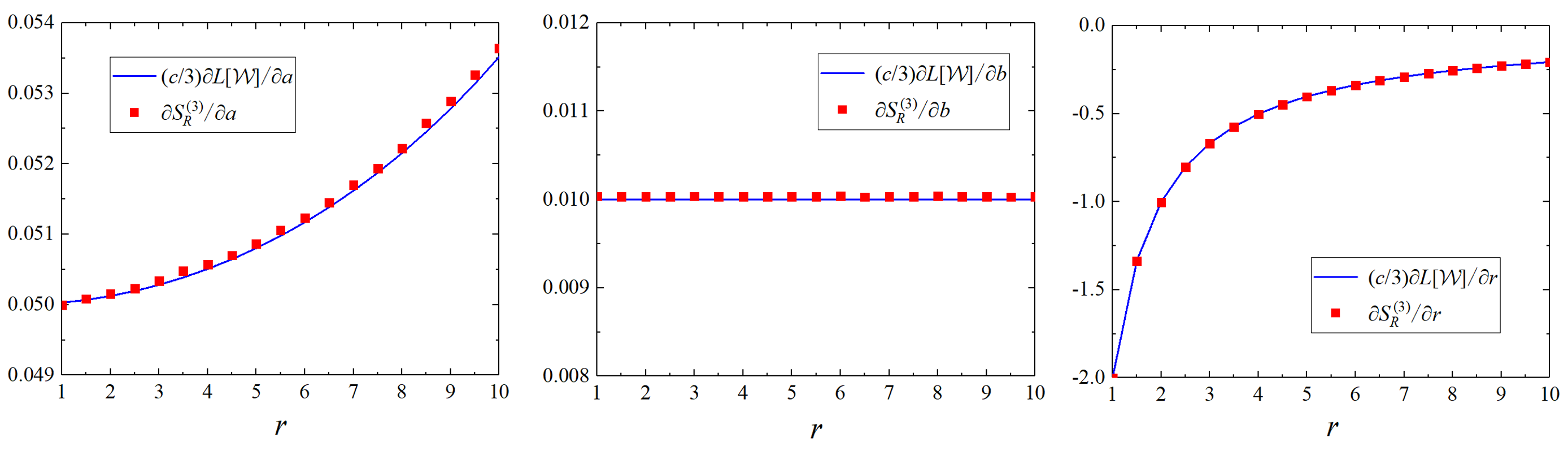}
	\caption{\label{fig-d-new}Comparison of the partial derivatives of $\frac{c}{3} L[\mathcal{W}]$ and $S_R^{(\mathtt{q} = 3)}$ (divided by $c/3$) with respect to $\xi = a,b,r$ at zero temperature. We take $a = 20$, $b = 100$, and $r$ ranging from $1$ to $10$. The holographic results and the field-theoretical results match well (the difference between them is within 0.4\%). }
\end{figure*}

\textit{Holographic calculation.}~The holographic dual of $S^{(3)}_R(A;B;C)$ is the minimal web denoted by the orange curves in Fig.~\ref{fig-conformalTrans-6pt}. Because of the $\mathbb{Z}_2$-symmetry about the $x = 0$ axis in Fig.~\ref{fig-conformalTrans-6pt}, the junction $I$ is expected to be on $x = 0$ and we denote it by $I (0,y_I)$. For general two points $(x_1, y_1)$ and $(x_2, y_2)$ on the Poincar\'e half plane, the distance between them is 
\begin{equation}
D \bigl((x_1, z_1), (y_2, y_2)\bigr) = \arccosh\frac{(x_1 - x_2)^2 + y_1^2 + y_2^2}{2 y_1 y_2}\ . 
\end{equation}
Therefore, the distance between $I$ and the RT surface of $c$ is given by 
\begin{equation}\label{eq-geo0}
   \min\nolimits_{\phi} D \bigl((0, y_I), (-a + r\cos\phi, r\sin\phi)\bigr)\ ,
\end{equation}
with $(-a + r\cos\phi, r\sin\phi)$ a point on the RT surface of $c$. The distances between $I$ and the RT surface of $a,b$ can be worked out similarly and we finally end up with the length of the bulk dual web
\begin{equation}\label{eq-LabrMinz}
\begin{split}
L\left[\mathcal{W}\right]&(a,b,r) = \min_{y_I} \Bigl[\log\left(b/y_I\right)\\
+& 2\arcsech \frac{2 y_I r}{\sqrt{\left( (a - r)^2 + y_I^2 \right) \left( (a + r)^2 + y_I^2 \right)}} \Bigr]\ ,
\end{split}
\end{equation}
where the first and second term in $[\cdots]$ comes from the distance between $I$ and the RT surface of $b$ and $\{a,c\}$. The minimization over $y_I$ in \eqref{eq-LabrMinz} will be evaluated numerically (see the blue curves in Fig.~\ref{fig-d-new}) when comparing with the field-theoretical results. 

\textit{Field-theoretical calculation.}~Now we do the calculation in CFT$_2$. Using the replica trick, we first represent the reflected multi-entropy as a path integral on a specific Riemann surface. 
Here, a new replica number $m$ is introduced for canonical purification (see Ref.~\cite{Dutta:2019gen} for more details about this replica trick). 

The $n$-th R\'enyi version of reflected multi-entropy is
\begin{equation}
S_R^{(\mathtt{q})}[n] = \frac{1}{1 - n} \frac{1}{n^{\mathtt{q} - 2}} \log\frac{\mathcal{Z}_{n^{\mathtt{q} - 1}, m}}{(\mathcal{Z}_{1, m})^{n^{\mathtt{q} - 1}}}\ ,
\end{equation}
and the reflected multi-entropy is obtained by taking the limit $n\to 1$ and $m\to 1$,
\begin{equation}\label{eq-RME}
S_R^{(\mathtt{q})} = \lim_{m \to 1}\lim_{n\to 1} \frac{1}{1 - n} \frac{1}{n^{\mathtt{q} - 2}} \log\frac{\mathcal{Z}_{n^{\mathtt{q} - 1}, m}}{(\mathcal{Z}_{1, m})^{n^{\mathtt{q} - 1}}}\ .
\end{equation}
Now we use the twist operator correlation function to express $\mathcal{Z}_{n^{\mathtt{q} - 1}, m}$, the path integral on a glued Euclidean spacetime with $m n^{\mathtt{q} - 1}$ copies. The conformal weights $h_i$ of twist operators $\sigma_i(x_i)$ can be read from the representation of cyclic groups~\cite{Lunin:2000yv},
\begin{equation}\label{eq-twistDim}
h_{i = 1, \dots, 2\mathtt{q}} = \frac{c}{24} \left( m - \frac{1}{m} \right) n^{\mathtt{q} - 1}\ . 
\end{equation}
We leave the derivation of the conformal weights~\eqref{eq-twistDim} and \eqref{eq-fusionDim} to Supplemental Material~\cite{SM}.
When $\mathtt{q} = 2$, one can check that \eqref{eq-twistDim} reduces to the conformal weight of the twist operators for the reflected entropy as we expected~\footnote{For reflected entropy, we have~\cite{Dutta:2019gen}
$$
    h_{g_A} = h_{g_A^{-1}} = \frac{cn(m^2 - 1)}{24 m}\ , 
$$
where $g_A$ and $g_A^{-1}$ are the twist operators located at the two endpoints of interval $A$.}. The conformal weights $h_a$ of the leading operators $\sigma_a$ in operator product expansion $\sigma_i(x_i)\sigma_j(x_j) \to \sigma_{a = ij}(x_a)$ are 
\begin{equation}\label{eq-fusionDim}
h_{a = ij} = \frac{c}{12} \left( n^{\mathtt{q} - 1} - n^{\mathtt{q} - 3}\right)\ ,
\end{equation}
which is twice the conformal weight of the twist operators for the original multi-entropy as we expected~\footnote{The conformal dimension of the multi-entropy twist operator is given in Eq.~(29) of~\cite{Harper:2024ker}. }.  It can also be checked that for $\mathtt{q} = 2$, \eqref{eq-fusionDim} reduces to twice the conformal weight of the twist operators for EE. 

Using these twist operators, $\mathcal{Z}_{n^{\mathtt{q} - 1}, m}$ can be represented by the six-point function
\begin{equation}\label{eq-6pt}
\mathcal{Z}_{n^{\mathtt{q} - 1}, m} = \left\langle \sigma_1 \sigma_2 \sigma_3 \sigma_4 \sigma_5 \sigma_6  \right\rangle_{\text{CFT}^{\otimes m n^{\mathtt{q} - 1}}}\ ,
\end{equation}
with $\sigma_i$ located at $x_i$. At large $c$ limit, the calculation of~\eqref{eq-6pt} can be converted to solving a monodromy problem~\cite{Belavin:1984vu, Zamolodchikov:1987avt}. Following the approach in~\cite{Hartman:2013mia} we compute the six-point function~\eqref{eq-6pt} at large $c$ limit~\cite{SM} and the partial derivatives of $S_R^{(3)}(A; B; C)$ with respect to $a,b,r$ are plotted in Fig.~\ref{fig-d-new}.

On the other hand, the holographic result is given by minimizing $L[\mathcal{W}](a,b,r)$ in \eqref{eq-LabrMinz}. The partial derivatives of $(c/3)L[\mathcal{W}](a,b,r)$ with respect to $a,b,r$ are plotted in Fig.~\ref{fig-d-new} and compared with the field-theoretical results. These two results match precisely. Notice that we are comparing the derivatives of the entropy. To match the entropy itself, we have to further include the contributions of the operator product expansion coefficients in the CFT calculation~\cite{SM}. 

We also examine the agreement between the field-theoretical and holographic results at finite temperature. From the CFT perspective, we compute the six-point function on an infinitely long cylinder with period $\beta_{\text{CFT}}$ while on the gravity side, the bulk dual web length is evaluated in the Ba\~nados-Teitelboim-Zanelli (BTZ) black hole background~\cite{Banados:1992wn}~\cite{SM}. We find that the reflected multi-entropy decreases with increasing temperature, indicating that thermal effects tend to wash out multipartite correlations. In the context of thermal systems, our proposal provides a concrete multipartite measure, whose potential connections to quantum chaos and information scrambling warrant further investigation. Another promising application of reflected multi-entropy is to explore whether it can serve as a diagnostic of topological order in mixed states~\cite{Ellison:2024svg, Wang:2023uoj}.

\subsection{Discussion}
In this Letter we introduced reflected multi-entropy as a multipartite measure for mixed state by combining the recent proposal of multi-entropy and canonical purification and found good agreement between the CFT results and the holographic results. One may worry about the potential bulk replica symmetry breaking for multi-entropy. While the bulk replica symmetry was assumed in the original holographic derivation~\cite{Gadde:2023zzj,Gadde:2022cqi}, recent work~\cite{Penington:2022dhr} shows this fails for $n > 2$~\footnote{For tripartite case, \cite{Gadde:2024taa} presents the following constraint
$$
\frac{1}{m_{AB}}+\frac{1}{m_{BC}}+\frac{1}{m_{CA}} > 1\ , \quad m_{AB,BC,CA}\in\mathbb{Z}_+\ , 
$$
with $m_{AB}$ the order of $\sigma_{A}^{-1}\sigma_{B}$. For the $n$-th R\'enyi multi-entropy ($m_{AB} = m_{BC} = m_{CA} = n$), this inequality is violated for $n > 2$, indicating replica symmetry breaking. }\nocite{Gadde:2024taa}. A similar issue arises for reflected multi-entropy with $n > 2$. Nevertheless, our holographic calculation focuses on the $n = 1$ case, evading the replica symmetry-breaking regime and ensuring the validity of our results.
Our findings strongly support a new class of duality between multipartite measures for mixed states and spacetime geometry. Moreover, there are two natural extensions of our work. First, our discussion can be generalized straightforward to multiple canonical purifications. 
Second, similar to the covariant generalizations of reflected entropy and multi-entropy~\cite{Dutta:2019gen,Gadde:2022cqi}, we expect that the ``covariant reflected multi-entropy'' is obtained as the area of the minimal extremal surface web rather than the global minimal one. 

\textit{Note added.} Recently, the authors in~\cite{Hayden:2023yij} found a special quantum state where the reflected entropy is not monotonically decreasing under partial trace thus fails to be a correlation measure. However, the reflected entropy remains a valid correlation measure for holographic states, as emphasized in the same paper. See also Ref.~\cite{Bueno:2020fle} for the monotonicity property of reflected entropy in free fields.

\textit{Acknowledgements.} We are grateful for useful discussions with our group members in Fudan University. We would like to thank Jinwei Chu for helpful discussions. This work is supported by NSFC grant 12375063. This work is also sponsored by Natural Science Foundation of Shanghai (21ZR1409800) as well as Shanghai Talent Development Fund.

\section{Supplemental Material}
\subsection{Derivation of conformal weights}\label{appx-ConformalDim}
In this section we derive the conformal weights of $\sigma_{i = 1,\dots,2\mathrm{q}}$ and $\sigma_{a = ij}$
\begin{equation}\label{eq-twistDim-SM}
h_{i = 1, \dots, 2\mathtt{q}} = \frac{c}{24} \left( m - \frac{1}{m} \right) n^{\mathtt{q} - 1}\ ,
\end{equation}
\begin{equation}\label{eq-fusionDim-SM}
h_{a = ij} = \frac{c}{12} \left( n^{\mathtt{q} - 1} - n^{\mathtt{q} - 3}\right)
\end{equation}
from the path integral representation of $\mathcal{Z}_{n^{\mathtt{q} - 1},m}$. We consider the $\{\mathtt{q} = 3, n = 3, m = 2\}$ case as an example. Replica trick in Fig.~\ref{fig-cycle} is determined by six twist operators $\sigma_{i = 1,\dots,6}$ located at the endpoints of the subsystems $A,B,C$ (see top left corner of Fig.~\ref{fig-purification}). 

These twist operators are made up of three nontrivial operators $\Sigma_{A,B,C}$ associated with the three subsystems $A,B,C$ respectively, which can be described by representation of cyclic groups. For instance, (123) in $\Sigma_A$ means that the lower edge of subsystem $A$ in the 1st copy is glued to the upper edge of $A$ in the 2nd copy, the lower edge of $A$ in the 2nd copy is glued to the upper edge of $A$ in the 3rd copy, and the lower edge of $A$ in the 3nd copy is glued to the upper edge of $A$ in the 1st copy. 

From Fig.~\ref{fig-cycle} we can read
\begin{widetext}
\begin{equation}\label{eq-SigmaABC}
\begin{split}
\Sigma_A &= (1,2) (3,4) (5,6) (7,8) (9,10) (11,12) (13,14) (15,16) (17, 18)\ ;\\
\Sigma_B &= (2,7) (8,13) (14,1) (4,9) (10,15) (16,3) (6,11) (12,17) (18,5)\ ;\\
\Sigma_C &= (2,3) (4,5) (6,1) (8,9) (10,11) (12,7) (14,15) (16,17) (18,13)\ .\\
\end{split}
\end{equation}
Then we have
\begin{equation}\label{eq-sigma123456}
\sigma_1 = \Sigma_A\ , \quad \sigma_2 = \Sigma_A^{-1}\ , \quad \sigma_3 = \Sigma_B\ , \quad \sigma_4 = \Sigma_B^{-1}\ , \quad \sigma_5 = \Sigma_C\ , \quad \sigma_6 = \Sigma_C^{-1}\ .
\end{equation}
Therefore, the conformal weights of $\sigma_i$~\footnote{The conformal weight of twist operator is given by~\cite{Lunin:2000yv}
$$
h = \frac{c}{24} \sum\nolimits_k p_k \left(k - \frac{1}{k}\right)\ ,
$$
with $p_k$ the number of $k$-cycles. } is $h_i = (c/24) \left(2 - 1/2\right) \times 9$, which is consistent with \eqref{eq-twistDim-SM}. For general $\{\mathtt{q},n,m\}$, since there are only $m$-cycles in $\sigma_{i = 1,\dots,2\mathrm{q}}$, their conformal weights $h_i$ is thus given by \eqref{eq-twistDim-SM}. 

\begin{figure*}[htbp]
	\centering
	\includegraphics[scale=0.9]{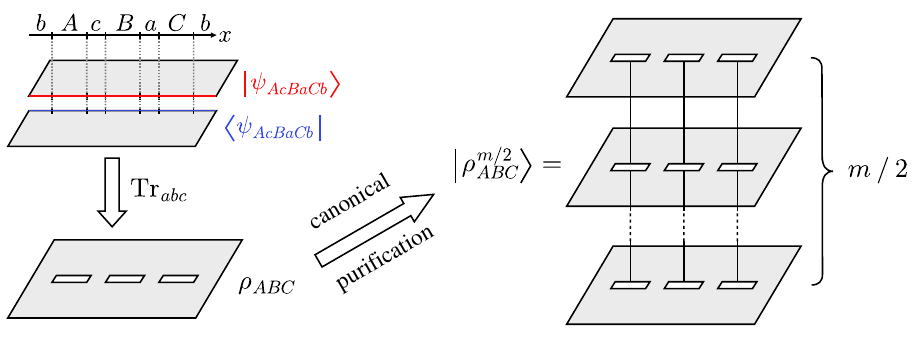}
	\caption{\label{fig-purification}Left: The path integral representation of the reduced density matrix $\rho_{ABC}$. Right: The path integral representation of $\ket{\rho_{ABC}^{m/2}}$, which is obtained from the canonical purification of $\rho_{ABC}$. Here $m$ should be even and it will be analytically continued to 1 in the end. }
\end{figure*}

\begin{figure*}[htbp]
    \centering
    \includegraphics[scale=0.9]{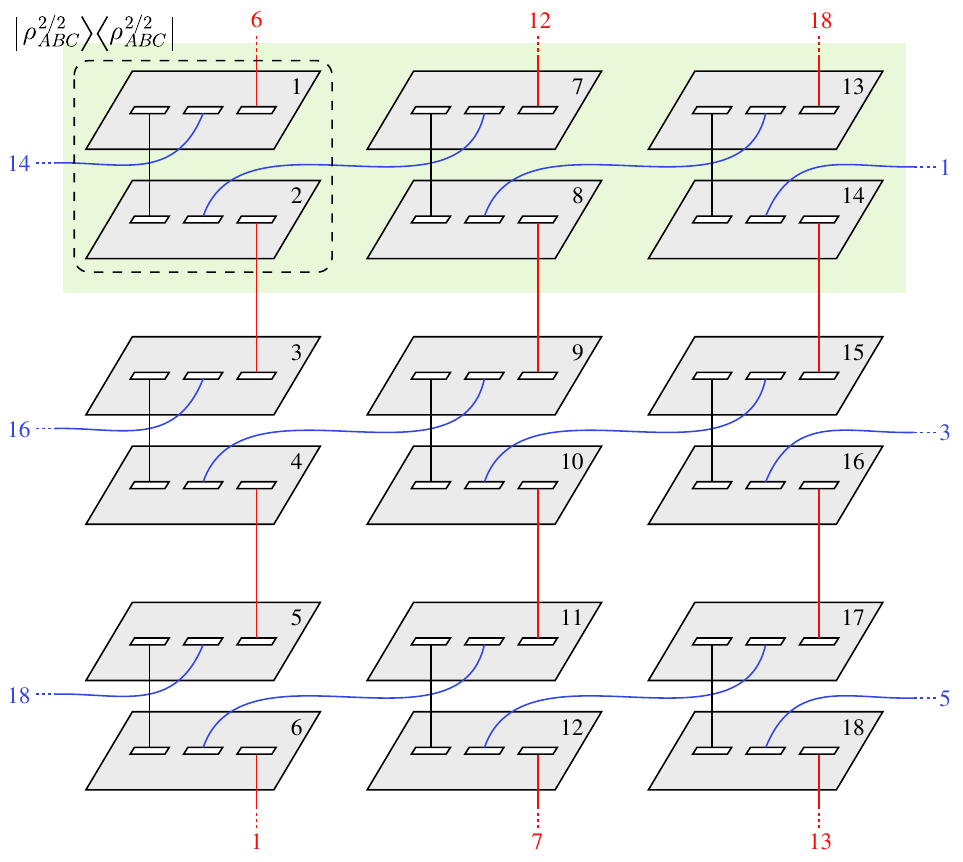}
    \caption{\label{fig-cycle}The path integral representation of $\mathcal{Z}_{n^{\mathtt{q} - 1}, m}$. Here we take $\mathtt{q} = 3$, $n = 3$ and $m = 2$. The part inside the dashed box is the density matrix $\ket{\rho_{ABC}^{2/2}}\bra{\rho_{ABC}^{2/2}}$. Some gluings involving replicas have not been shown. In this figure, if the lower edge of a subsystem in the i-th copy is glued to the upper edge in the j-th copy, then the upper edge of the same subsystem in the i-th copy is also glued to the lower edge in the j-th copy.} 
\end{figure*}

From \eqref{eq-SigmaABC} and \eqref{eq-sigma123456}, we can obtain leading operator $\sigma_{a = 16,23,45}$ in the OPE of $\sigma_i$ and $\sigma_j$
\begin{equation}\label{eq-twistFusion}
\begin{split}
\sigma_{16} &= \sigma_1\sigma_6 = \Sigma_A \Sigma_C^{-1} = (1,5,3) (2,4,6) (7,11,9) (8,10,12) (13,17,15) (14,16,18)\ ;\\
\sigma_{23} &= \sigma_2\sigma_3 = \Sigma_A^{-1} \Sigma_B = (1,13,7) (2,8,14) (3,15,9) (4,10,16) (5,17,11) (6,12,18)\ ;\\
\sigma_{45} &= \sigma_4\sigma_5 = \Sigma_B^{-1} \Sigma_C = (1,11,15) (2,16,12) (3,7,17) (4,18,8) (5,9,13) (6,14,10)\ .\\
\end{split}
\end{equation}
Therefore, the conformal weights of $\sigma_a$ are $h_a = (c/24) \left(3 - 1/3\right) \times 6$, which are consistent with \eqref{eq-fusionDim-SM}. Now we generalize this result to general $\left\{\mathtt{q}, n, m\right\}$ case, where we have $m n^{\mathtt{q - 1}}$ copies and they form a $\underbrace{n\times n\times \cdots \times n}_{\mathtt{q} -1}$ lattice with $m$ copies at each cell. We focus on the row containing the first cell~\footnote{This row contains $nm$ replicas in total. For $n = 3$ and $m = 2$ case, this row corresponds to the green shaded part in Fig.~\ref{fig-cycle}. } and consider the product of the parts of $\Sigma_A^{-1}$ and $\Sigma_B$ restricted to the cells in this row. We use $\tilde{\Sigma}_A$ and $\tilde{\Sigma}_B$ to denote the parts of $\Sigma_A$ and $\Sigma_B$ restricted to this row, and they are given by (since $m$ is even, here we take $m = 2\ell$)
\begin{equation}\label{eq-cycleGeneral}
\begin{split}
\tilde{\Sigma}_A =\ 
&(1, 2, \dots, \ell, {\color{purple}\ell + 1, \ell + 2, \dots, 2\ell}) ({\color{blue}nm + 1, nm + 2, \dots, nm + \ell}, nm + \ell + 1, nm + \ell + 2, \dots, nm + 2\ell)\\
&(2nm + 1, \dots, 2nm + 2\ell) \cdots ((n - 1)nm + 1, \dots, (n - 1)nm + 2\ell)\ ,\\
\tilde{\Sigma}_B =\ 
&({\color{purple}\ell + 1, \ell + 2, \dots, 2\ell}, {\color{blue}nm + 1, nm + 2, \dots, nm + \ell}) (nm + \ell + 1, \dots, nm + 2\ell, 2nm + 1, \dots, 2nm + \ell)\cdots\\
&((n - 2)nm + \ell + 1, \dots, (n - 2)nm + 2\ell, (n - 1)nm + 1, \dots (n - 1)nm + \ell)\\
&((n - 1)nm + \ell + 1, \dots, (n - 1)nm + 2\ell, 1, \dots, \ell)\ .
\end{split}
\end{equation}
The product $\tilde{\Sigma}_A^{-1} \tilde{\Sigma}_B$ is 
\begin{equation}\label{eq-fusionGeneral}
\tilde{\Sigma}_A^{-1} \tilde{\Sigma}_B = \left(\ell, (n - 1)nm + \ell, (n - 2)nm + \ell, \dots, nm + \ell \right ) \left(2\ell, nm + 2\ell, 2nm + 2\ell, \dots, (n - 1)nm + 2\ell \right) \ , 
\end{equation}
\end{widetext}
which contains two $n$-cycles. One can check that when $n = 3, m = 2$, \eqref{eq-cycleGeneral} and \eqref{eq-fusionGeneral} reduces to (part of) \eqref{eq-SigmaABC} and \eqref{eq-twistFusion}. Note that our analysis here is restricted to a single row. Since there are $n^{\mathtt{q} - 2}$ rows in total, the complete $\Sigma_A^{-1} \Sigma_B$ operator consists of $2 n^{\mathtt{q} -2}$ $n$-cycles, resulting in the conformal weight~\eqref{eq-fusionDim-SM}. 
\subsection{CFT computation of the six-point function}\label{appx-6pt}
The reflected multi-entropy can be calculated by replica trick
\begin{equation}\label{eq-RME-SM}
S_R^{(\mathtt{q})} = \lim_{m \to 1}\lim_{n\to 1} \frac{1}{1 - n} \frac{1}{n^{\mathtt{q} - 2}} \log\frac{\mathcal{Z}_{n^{\mathtt{q} - 1}, m}}{(\mathcal{Z}_{1, m})^{n^{\mathtt{q} - 1}}}\ ,
\end{equation}
with $\mathcal{Z}_{n^{\mathtt{q} - 1}, m}$ the partition function on a glued Euclidean spacetime with $m n^{\mathtt{q} - 1}$ copies, which can be expressed using the twist operator correlation function. For the $\mathrm{q} = 3$ case we are considering, it is given by a six-point function 
\begin{equation}\label{eq-6pt-SM}
\mathcal{Z}_{n^{3 - 1}, m} = \left\langle \sigma_1 \sigma_2 \sigma_3 \sigma_4 \sigma_5 \sigma_6  \right\rangle_{\text{CFT}^{\otimes m n^{3 - 1}}}\ ,
\end{equation}
with the six points located at $x_i$ 
\begin{equation}\label{SMeq-coorpara}
\begin{split}
x_6 &= -x_1 = b\ , \\
x_5 &= -x_2 = a+r\ , \\
x_4 &= -x_3 = a-r
\end{split}
\end{equation}
and the relevant conformal weights given by \eqref{eq-twistDim-SM} and \eqref{eq-fusionDim-SM}. In this section we give a numerical calculation of the six-point function \eqref{eq-6pt-SM}. Utilizing the conformal transformation
\begin{equation}
x_i \to z_i = \frac{(x_i - x_2)(x_1 - x_6)}{(x_2 - x_6)(x_i - x_1)}\ ,
\end{equation}
$\{x_1, x_2, x_6\}$ are mapped to $\{z_1, z_2, z_6\} = \{\infty, 0, 1\}$ while $\{x_3, x_4, x_5\}$ are sent to $\{z_3, z_4, z_5\}$. At the large $c$ limit with $h_{i = 1,\dots,6} / c$ and $h_{a = 16,23,45} / c$ fixed (here $h_i$ denotes the conformal weight of the external operator $\sigma_i(x_i)$ and $h_a$ the weight of the internal operator), the six point function 
\begin{equation}\label{eq-sixPt1}
\left\langle \sigma_1(z_1) \sigma_2(z_2) \sigma_3(z_3) \sigma_4(z_4) \sigma_5(z_5) \sigma_6(z_6)  \right\rangle
\end{equation}
can be approximated by
\begin{equation}\label{eq-block6ptdo}
\begin{split}
&\left\langle \sigma_1(z_1) \sigma_2(z_2) \sigma_3(z_3) \sigma_4(z_4) \sigma_5(z_5) \sigma_6(z_6)  \right\rangle\\
\approx\ &c_{1,6}^{16} c_{2,3}^{23} c_{4,5}^{45} c_{16,23}^{45} \mathcal{F}(z_i) \mathcal{F}(\bar{z}_i)\ , 
\end{split}
\end{equation}
where $a = 16,23,45$ labels the leading order primary with conformal wight~\eqref{eq-fusionDim-SM}, $c_{i,j}^{a}$ is the OPE coefficient, and $\mathcal{F}$ is the six-point Virasoro block, which exponentiates at large $c$ limit~\cite{Hartman:2013mia}
\begin{equation}
\mathcal{F} \sim \exp\left[ -\frac{c}{6} f\left(\frac{h_a}{c}, \frac{h_i}{c}, z_i\right)\right]\ .  
\end{equation}
Here $f$ is called the semi-classical block, which can be obtained by solving the following monodromy problem. Consider the second-order differential equation
\begin{equation}\label{eq-DiffEq}
\psi''(z) + T(z) \psi(z) = 0\ , 
\end{equation}
where $T(z)$ is given by
\begin{equation}
T(z) = \sum_{i = 1}^{6} \left(\frac{6h_i / c}{(z - z_i)^2} - \frac{c_i}{z - z_i}\right)\ , 
\end{equation}
with $c_i$ the accessory parameters satisfying
\begin{equation}\label{eq-ci-cond}
\begin{split}
&\sum\nolimits_{i = 1}^{6} c_i  = 0 \ , \\
&\sum\nolimits_{i = 1}^{6} \left(c_i z_i - 6 h_i/c\right) = 0\ , \\
&\sum\nolimits_{i = 1}^{6} \left(c_i z_i^2 - 12 z_i  h_i/c\right) = 0\ , 
\end{split}
\end{equation}
which guarantee that $T(z)$ vanishes as $z^{-4}$ at infinity. As a second-order differential equation, \eqref{eq-DiffEq} has two solutions $\psi_1$ and $\psi_2$. Taking these solutions on a closed contour around some singular point, they will undergo some monodromy
\begin{equation}\label{eq-monodromyM}
\begin{pmatrix}
\psi _{1}\\
\psi _{2}
\end{pmatrix}\rightarrow M\begin{pmatrix}
\psi _{1}\\
\psi _{2}
\end{pmatrix}\ . 
\end{equation}

The accessory parameters $c_i$ can be determined by \eqref{eq-ci-cond} and the following three equations 
\begin{equation}
\text{Tr} M_a = -2\cos\left( \pi \sqrt{1 - \frac{24}{c}h_a} \right)\ , \quad a = 16,23,45, 
\end{equation}
where $M_{a = 16}$ denotes the $2\times 2$ monodromy matrix for the cycle $\gamma_{16}$ enclosing $z_1$ and $z_6$, and similarly for $a = 23,45$; 
\begin{figure}[htbp]
    \centering
    \includegraphics[scale=0.8]{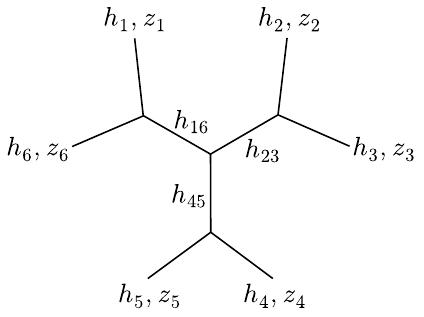}
    \caption{Dominant fusion channel of the 6-point function~\eqref{eq-sixPt1}. }
    \label{fig-fusion}
\end{figure}
$h_{16}$ is the conformal weight of the leading operator in the OPE contraction of $\sigma_1(z_1)$ and $\sigma_6(z_6)$ as shown in Fig.~\ref{fig-fusion}, similarly for $a = 23,45$. 
From \eqref{eq-fusionDim-SM} we know that $h_{16} = h_{23} = h_{45} = \frac{c}{12}\left(n^2 - 1\right)$. 

The relation between the semi-classical block $f$ and the accessory parameters $c_i$ is given by
\begin{equation}
\partial f / \partial z_i = c_i\ . 
\end{equation}
Therefore, we can calculate the partial derivative of $S_R^{(\mathtt{q})}$ with respect to the coordinate parameters $\xi = a,b,r$ in \eqref{SMeq-coorpara}. From \eqref{eq-RME-SM}, \eqref{eq-6pt-SM} and \eqref{eq-block6ptdo} we obtain
\begin{equation}\label{eq-6pt-derivative}
\begin{split}
\frac{\partial S_R^{(\mathtt{q})}}{\partial \xi} 
&= \lim_{n \to 1} \frac{1}{n - 1} \frac{1}{n^{3 - 2}} \frac{c}{3} \sum_{i = 1}^{6} \frac{\partial f}{\partial z_i} \frac{\partial z_i}{\partial \xi}\\
&= \lim_{n \to 1} \frac{1}{n - 1} \frac{1}{n^{3 - 2}} \frac{c}{3} \sum_{i = 1}^{6} c_i \frac{\partial z_i}{\partial \xi}\ . 
\end{split}
\end{equation}
\subsection{Reflected multi-entropy at finite temperature}
In this section we consider reflected multi-entropy in CFT at finite temperature, the bulk dual of which is the BTZ black hole~\cite{Banados:1992wn}
\begin{equation}\label{eq-BTZ}
\mathrm{d}s^2 = \frac{1}{y^2} \left(f(y)\mathrm{d}\tau^2 + \frac{\mathrm{d}y^2}{f(y)} + \mathrm{d}x^2  \right)\ ,
\end{equation}
where $f(y) = 1 - (y/y_H)^2$ with $y = y_H$ the horizon as shown in Fig.~\ref{fig-finiteT-holo}, $\tau \sim \tau + 2\pi y_H$, $\beta_{\text{CFT}} = 2\pi y_H$, and we have set the AdS radius to be 1. To be specific, we consider the reflected multi-entropy among the following three intervals
\begin{equation}
A = [-b, -a]\ ,\quad B = [-a, a]\ ,\quad C = [a, b]\ , 
\end{equation}
on the time slice $\tau = 0$ as illustrated in Fig.~\ref{fig-finiteT-holo}. To simplify, we consider the case that $b$ goes to infinity.
\subsubsection{Holographic calculation.}
\begin{figure}[tbp]
	\centering
	\includegraphics[width=0.9\columnwidth]{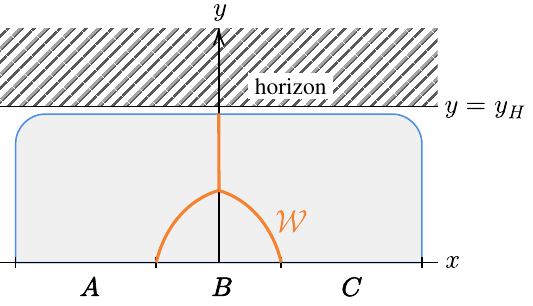}
	\caption{\label{fig-finiteT-holo} The holographic dual of $S_R^{(\mathtt{q} = 3)}(A; B; C)$ at finite temperature. }
\end{figure}

Now we compute the holographic reflected multi-entropy in the presence of BTZ black hole. The distance between two arbitrary points $(\tau, x, y)$ and $(\tau', x', y')$ in the BTZ metric \eqref{eq-BTZ} is given by
\begin{equation}
\begin{split}
&\tilde{D}\bigl((\tau, x, y),(\tau', x', y')\bigr)\\
= &\arccosh \left(T_1 T'_1 + T_2 T'_2 -X_1 X'_1 -X_2 X'_2\right)\ ,
\end{split}
\end{equation}
with
\begin{equation}
\begin{split}
T_1 &= \sqrt{\frac{y_H^2}{y^2} - 1}\sinh\frac{\tau}{y_H}\ ,\quad T_2 = \frac{y_H}{y}\cosh\frac{x}{y_H}\ ,\\
X_1 &= \sqrt{\frac{y_H^2}{y^2} - 1}\cosh\frac{\tau}{y_H}\ ,\quad X_2 = \frac{y_H}{y}\sinh\frac{x}{y_H}\ .
\end{split}
\end{equation}
Due to the $\mathbb{Z}_2$-symmetry about the $x = 0$ axis in Fig.~\ref{fig-finiteT-holo}, we denote the junction point $I$ by $I (0,y_I)$. Since we send $b$ to $\infty$, the RT surface of $A \cup B \cup C$ will be very close to the horizon. Therefore, the sum of the total three geodesics is given by
\begin{equation}\label{eq-T-geo}
2 \tilde{D}\bigl((0,a,\epsilon),(0,0,y_I)\bigr) + \tilde{D}\bigl((0,0,y_H),(0,0,y_I)\bigr)\ ,
\end{equation}
with $\epsilon$ the UV cutoff. Minimizing \eqref{eq-T-geo} with respect to $y_I$, we end up with $y_I = \frac{\sqrt{3} y_H \sinh(a/y_H)}{2\cosh(a/y_H) - 1}$ and
\begin{equation}\label{eq-T-holoResult}
\begin{split}
&L[\mathcal{W}](a) = \arccosh\frac{2\cosh\frac{a}{y_H} - 1}{\sqrt{3} \sinh\frac{a}{y_H}}\\
&+ 2\log\frac{2 y_H \left(2\cosh^2\frac{a}{y_H} - |\cosh\frac{a}{y_H} - 2| - \cosh\frac{a}{y_H}\right)}{\sqrt{3} \epsilon \sinh\frac{a}{y_H}}\ .
\end{split}
\end{equation}
\subsubsection{Field-theoretical calculation.}
\begin{figure}[htbp]
	\centering
	\includegraphics[scale=0.8]{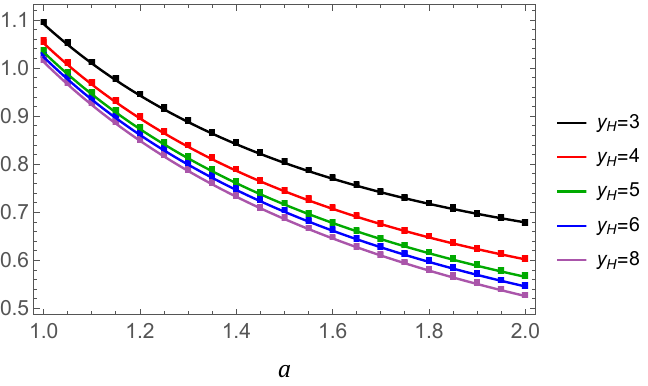}
	\caption{\label{fig-T-compare}The partial derivatives of $\frac{c}{3} L[\mathcal{W}]$ and $S_R^{(\mathtt{q} = 3)}$ (divided by $c/3$) with respect to $a$ at different temperature are represented by lines and dots respectively. }
\end{figure}

\begin{figure}[htbp]
	\centering
	\includegraphics[scale=0.8]{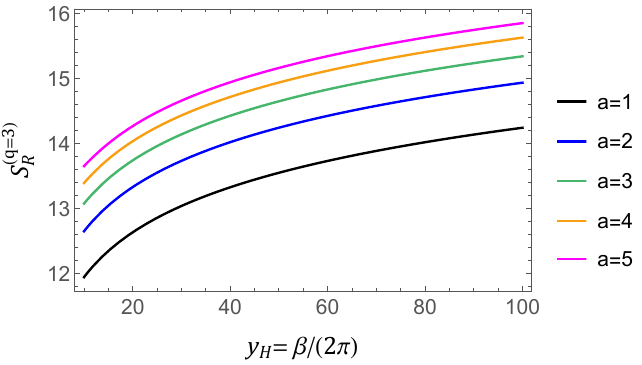}
	\caption{\label{fig-SR-beta}The variation of reflected multi-entropy $S_R^{(\mathtt{q} = 3)}$ (divided by $c/3$) with respect to $y_H$ for different $a$. }
\end{figure}

From the CFT perspective, we need $\mathcal{Z}_{n^{\mathtt{q} - 1},m}$, which can be obtained from the six-point function on a cylinder of infinite length and period $\beta_{\text{CFT}}$
\begin{equation}\label{eq-T-6pt}
\left<\sigma_1 \sigma_2 \sigma_3 \sigma_4 \sigma_5 \sigma_6\right>_\text{cylinder}\ ,
\end{equation}
with $\sigma_i$ located at $\tau_i = 0$ and $x = x_i$, which is given by
\begin{equation}
\begin{split}
x_6 &= -x_1 = b = \infty\ ,\\
x_5 &= -x_2 = a + \epsilon\ ,\\
x_4 &= -x_3 = a - \epsilon\ .
\end{split}
\end{equation}
The conformal weight of $\sigma_i$ is given by \eqref{eq-twistDim-SM} as before. To compute \eqref{eq-T-6pt} we perform the conformal transformation (here $w_i = x_i + \mathtt{i}\tau_i$)
\begin{equation}
z_i = 1 - \exp\frac{2\pi(w_2 - w_i)}{\beta_{\text{CFT}}}\ ,
\end{equation}
which sends $\{w_1, w_2, w_6\}$ to $\{\infty, 0, 1\}$ and $\{w_3, w_4, w_5\}$ to $\{z_3, z_4, z_5\}$. Following the same method as at zero temperature, we can numerically evaluate the derivative of the reflected multi-entropy with respect to $a$ and compare it with the holographic result~\eqref{eq-T-holoResult} as plotted in Fig.~\ref{fig-T-compare}. Again these two results match precisely.
The variation of reflected multi-entropy with respect to $y_H = \beta/(2\pi)$ for different interval length $a$ is shown in Fig.~\ref{fig-SR-beta}. As the temperature rises, the reflected multi-entropy decreases, which fits our expectations since in thermal CFT, the reflected entropy has similar behavior in the phase where the entanglement wedge cross section ends on the horizon (see Eq.~(24) of \cite{Takayanagi:2017knl} for example). As the parameter $a$ (which characterizes the size of region  $B$) increases, $S_R^{(3)}$ increases, indicating that it is a correlation measure for holographic states. 
\subsection{The OPE coefficient}\label{appx-OPE}
The monodromy method can only determine the derivatives of the six-point function~\eqref{eq-6pt-derivative}. To obtain the full field-theoretical results, we need to fix the OPE coefficients appearing in~\eqref{eq-block6ptdo}. There are two kinds of OPE coefficients, $c_{i,j}^{a}$ and $c_{a,b}^{c}$, with $i,j,\dots = 1,\dots,6$ labeling the twist operators with conformal weights~\eqref{eq-twistDim-SM} and $a,b,c = 16,23,45$ labeling those with conformal weights~\eqref{eq-fusionDim-SM}. The OPE coefficient $c_{i,j}^{a}$ is similar to the OPE coefficient of reflected entropy, while $c_{a,b}^{c}$ is the square of the OPE coefficient of multi-entropy. Let us first focus on $c_{i = 2,j = 3}^{a = 23}$, which corresponds to the fusion $\sigma_2 \sigma_3 \to \sigma_{23}$. We consider
\begin{equation}
\sigma_2 = \Sigma_A^{-1}\ , \quad \sigma_3 = \Sigma_B\ , \quad \sigma_{23} = \Sigma_A \Sigma_B^{-1}\ .
\end{equation}
The three point function $\langle \sigma_2 \sigma_3 \sigma_{23} \rangle$ computes the partition function of the replica space, a $m n^{\mathtt{q} - 1}$-sheet Riemann surface, which in fact factorizes into $n^{\mathtt{q - 2}}$ number of $m n$-sheet Riemann surfaces. The $m n$-sheet Riemann surface is nothing but the replica geometry appearing in the replica trick of reflected entropy (see appendix C of Ref.~\cite{Dutta:2019gen}), thus we have
\begin{equation}
\langle \sigma_2 \sigma_3 \sigma_{23} \rangle = \langle \sigma_{g_A^{-1}} \sigma_{g_B} \sigma_{g_A g_B^{-1}} \rangle^{n^{\mathtt{q} - 2}}\ ,
\end{equation}
where $g_A$ and $g_B$ are the twist operators for reflected entropy. Therefore, $c_{2,3}^{23}$ should be the $n^{\mathtt{q} - 2}$-th power of the reflected entropy OPE coefficient $C_{n,m}$, 
\begin{equation}\label{eq-OPE-1}
\begin{split}
&c_{2,3}^{23} = \left(C_{n,m}\right)^{n^{\mathtt{q} - 2}} = (2 m)^{-4 h_{n, \mathtt{q}}}\ , \\
&\text{with}\ h_{n, \mathtt{q}} = \frac{c}{24} \left( n^{\mathtt{q} - 1} - n^{\mathtt{q} - 3}\right)\ ,
\end{split}
\end{equation}
which reduces to the OPE coefficient of reflected entropy when $\mathtt{q} = 2$. 

The other OPE coefficient, $c_{a = 16,b = 23}^{c = 45}$, is the square of the multi-entropy OPE coefficient $C_n$ (since in our case the corresponding conformal weights double), whose $n$-derivative can been fixed from holography~\cite{Gadde:2023zzj, Harper:2024ker}
\begin{equation}\label{eq-OPE-2}
-\partial_n \log(c_{a,b}^{c})\big|_{n = 1} = -2\partial_n \log(C_n)\big|_{n = 1} = c \log\left( \frac{2}{\sqrt{3}} \right)\ .
\end{equation}
With the result~\eqref{eq-OPE-1} and the assumption~\eqref{eq-OPE-2}, the six-point function can be completely fixed as following: in the adjacent limit 
\begin{equation}
|x_1 - x_6| = |x_2 - x_3| = |x_4 - x_5| = 2\epsilon \to 0\ ,
\end{equation}
the six-point function reduces to a three-point function
\begin{equation}
\begin{split}
\left\langle \sigma_1 \sigma_2 \sigma_3 \sigma_4 \sigma_5 \sigma_6  \right\rangle 
&\to c_{1,6}^{16}c_{2,3}^{23}c_{4,5}^{45}\left\langle \sigma_{a = 16} \sigma_{b = 23} \sigma_{c = 45} \right\rangle\\
&= (2 m)^{-12 h_{n, \mathtt{q} = 3}} \left\langle \sigma_{a} \sigma_{b} \sigma_{c} \right\rangle
\end{split}
\end{equation}
with $\sigma_{a = 16}$ located at $x_a = (x_1 + x_6)/2$ and similar for $\sigma_b$, $\sigma_c$. 
The three-point function is given by
\begin{widetext}
\begin{equation}
\left\langle \sigma_{a} \sigma_{b} \sigma_{c} \right\rangle = \frac{c_{a,b}^{c}}{\left(x_{ab}/(2\epsilon)\right)^{2h_a + 2h_b - 2h_c} \left(x_{bc}/(2\epsilon)\right)^{2h_b + 2h_c - 2h_a} \left(x_{ca}/(2\epsilon)\right)^{2h_c + 2h_a - 2h_b}} = c_{a,b}^{c}\left[\frac{(2\epsilon)^3}{x_{ab} x_{bc} x_{ca}}\right]^{\frac{c}{6}(n^2 - 1)}\ ,
\end{equation}
where $x_{ab} \equiv |x_a - x_b|$ and we take the cutoff to be $2\epsilon$. With the adjacent limit result as an input, we can further use the monodromy method to determine the value of six-point function in all the parameter region (since the monodromy method determines the derivatives of the six-point function). The consistency between the field-theoretical and holographic results can also be verified in the adjacent limit. The field-theoretical result is given by
\begin{equation}
\begin{split}
S_R^{(3)}[n] 
&= \frac{1}{1-n}\frac{1}{n}\log \left\langle \sigma_1 \sigma_2 \sigma_3 \sigma_4 \sigma_5 \sigma_6  \right\rangle\\
&= -12 \frac{1}{1-n}\frac{1}{n} h _{n,\mathtt{q} = 3} \log(2 m) - \frac{c}{6}\frac{1}{1-n}\frac{1}{n}\left(n^2 - 1\right)\log\frac{x_{ab} x_{bc} x_{ca}}{(2\epsilon)^3} + \frac{1}{1-n}\frac{1}{n} \log c_{a,b}^{c}\\
&= \frac{c}{6}\left(1 + \frac{1}{n}\right)\log\frac{x_{ab} x_{bc} x_{ca}}{m^3\epsilon^3} + \frac{1}{1-n}\frac{1}{n}\log c_{a,b}^{c}\ .
\end{split}
\end{equation}
\end{widetext}
In the $m,n\to 1$ limit, 
\begin{equation}
\begin{split}
S_R^{(3)}
&= \frac{c}{3}\log\frac{x_{ab} x_{bc} x_{ca}}{\epsilon^3} - \partial_n \log(c_{a,b}^{c})\big|_{n = 1}\\
&= \frac{c}{3}\log\frac{x_{ab} x_{bc} x_{ca}}{\epsilon^3} + c \log\left( \frac{2}{\sqrt{3}} \right)\ ,
\end{split}
\end{equation}
which is the same as $(c/3)L[\mathcal{W}]$ (see Ref.~\cite{Harper:2024ker} for the calculation of $L[\mathcal{W}]$ in the adjacent limit).

\bibliography{biblio}

\end{document}